\documentclass[aps,prd,preprint,superscriptaddress,nofootinbib,showpacs]{revtex4}
\usepackage{graphics}
\usepackage{epsfig}
\usepackage{latexsym}
\usepackage{colordvi}
\usepackage{amsmath}
\usepackage{amssymb}

\begin{document}
\preprint{\parbox[b]{1in}{ \hbox{\tt PNUTP-06/A01} \hbox{\tt
KIAS-P06002} }}

\title{Holographic Estimate of Oblique Corrections for Technicolor }

\author{Deog Ki Hong}
\email[E-mail: ]{dkhong@pusan.ac.kr} \affiliation{Department of
Physics, Pusan National University,
             Busan 609-735, Korea}

\author{Ho-Ung Yee}
\email[E-mail: ]{ho-ung.yee@kias.re.kr} %\\
\affiliation{School of Physics, Korea Institute for Advanced Study,
Seoul 130-012, Korea}

\vspace{0.1in}

\date{\today}

\begin{abstract}
We study the oblique corrections to the electroweak interaction in
the holographic model of technicolor theories. The oblique S
parameter is expressed in terms of a solution to the equations of
motion for the AdS bulk gauge fields. By analyzing the solution, we
establish a rigorous proof that the S parameter is positive and is
reduced by walking. We also present the precise numerical values for
the S parameter of various technicolor models by solving the
equations numerically.
\end{abstract}
%%% 11.25.Tq Gauge/string duality
%%% 11.10.Kk Field theory in dimension other than four
%%% 11.25.Wx String and brane phenomenology
%%% 12.38.Cy Summation of perturbation theory in QCD
%%% 12.60.Fr Extension of electroweak Higgs sector
%%% 12.15.Ji Applications of electroweak models to specific processes
%%% 12.15.Lk Electroweak radiative corrections
%%% 12.60.Nz Technicolor models

\maketitle

\newpage

%\section{Introduction}
The development of the  anti-de Sitter/conformal field theory
(AdS/CFT) correspondence~\cite{Maldacena:1997re} has attracted a lot
of interest in recent years, since it may shed some light on
strongly coupled gauge theories, whose understanding has been
otherwise quite limited. Inspired by recent
success~\cite{Erlich:2005qh,DaRold:2005zs} in understanding the
infrared (IR) dynamics of quantum chromodynamics (QCD) with its
holographic dual, we apply the AdS/CFT correspondence to estimate
the oblique corrections of technicolor theories.

The precision electroweak data has shown that the standard model of
electroweak interaction is extremely stable radiatively, which is
often expressed in terms of vanishingly small Peskin-Takeuchi S, T,
U parameters~\cite{Peskin:1990zt,Holdom:1990tc}. Nonzero values of
those parameters indicate new physics beyond the standard model. Any
model for the new physics is highly constrained by the measured
values. Since the oblique correction of technicolor is dominated by
the nonperturbative dynamics  in the IR region, its precise estimate
has been a major hurdle for technicolor theories to be candidates
for physics beyond the standard model. Recently, however, a new
class of technicolor models
(techni-orientifold)~\cite{Sannino:2004qp,Hong:2004td}, which uses a
higher-dimensional representation for techniquarks and the
correspondence with ${\cal N}=1$ super Yang-Mills theory at large
$N$~\cite{Armoni:2003gp}, has been proposed. The models need only
small number of technifermions to be in the conformal window, free
from the flavor-changing neutral-current problem, and thus have
naturally small oblique
corrections~\cite{Hong:2004td,Dietrich:2005jn,Evans:2005pu}.

In this article we attempt to calculate precisely the S parameter
for technicolor theories, using the AdS/CFT correspondence. (Similar
attempts were made in 5-dimensional Higgsless
models~\cite{Csaki:2003zu}.) We also show rigorously that the S
parameter is strictly positive for the holographic dual models of
dynamical electroweak symmetry breaking and the walking behavior of
the technicolor dynamics reduces the S parameter substantially.

%\section{holographic dual}
According to AdS/CFT correspondence, to every operators in CFT
there correspond AdS bulk fields. The bulk fields that satisfy the
equations of motion are the sources for the CFT operators, when
evaluated at the ultraviolet (UV) boundary of the AdS space, and
their  action at the UV boundary is nothing but the generating
functional for the connected Green functions of those operators in
CFT.

For a near conformal theory whose IR scale is generated far below
the UV scale, we may take a slice of the AdS metric as
\begin{equation}
ds^2=\frac{1}{z^2}\left(-dz^2+\eta^{\mu\nu}dx_{\mu}dx_{\nu}\right),
\quad \epsilon\le z\le z_m\,,
\end{equation}
where $z=\epsilon$ ($z=z_m$) is the UV (IR) boundary, and
$\eta^{\mu\nu}$ is the four-dimensional Minkowski metric. For the
boundary conformal field theory we choose a strongly coupled
technicolor theory with ${\rm SU}(N)_{TC}$ gauge group and $N_{TF}$
massless techniquarks, $q^{\alpha}$ ($\alpha=1,\cdots, N_{TF}$),
which may be fundamental under the technicolor gauge group as in the
walking technicolor~\cite{Holdom:1984sk} or second rank tensor as in
the techni-orientifold~\cite{Sannino:2004qp,Hong:2004td}. Since the
technicolor model we choose for the boundary field theory is not
exactly conformal, the AdS/CFT correspondence does not hold in a
strict sense. We expect however the holographic dual description of
technicolor to work reasonably well, as the conformal symmetry is
mildly broken, especially for the technicolor theories in the conformal window. The
boundary operators we are interested in are the scalar, $\bar
q_{\alpha}q^{\beta}$, and the (axial) vector current,
$J_{V(A)}^{a\mu}=\bar q \gamma^{\mu}(\gamma_5)t^a\,q$, of
techniquarks. ($t^a$'s are the ${\rm SU}(N_{TF})$ generators,
normalized as ${\rm Tr}\,t^at^b=1/2\,\delta^{ab}$.) The
corresponding bulk action is then given as
\begin{equation}
S=\int d^5x\sqrt{g}\,{\rm
Tr}\left[\left|DX\right|^2-m_5^2\left|X\right|^2-
\frac{1}{2g_5^2}\left(F_L^2+F_R^2\right)\right]
\end{equation}
where $m_5$ is the mass of the bulk scalar field $X$ and the
covariant derivative
$D_{\mu}X=\partial_{\mu}X-i{A_{L}}_{\mu}\,X+iX\,{A_{R}}_{\mu}$. The
bulk mass is determined by the
relation~\cite{Gubser:1998bc,Witten:1998qj},
$\Delta\,(\Delta-4)=m_5^2$, where $\Delta$ is the dimension of the
corresponding boundary operator $\bar q_{\alpha}q^{\beta}$. $F_L$
and $F_R$ are the field strength tensors of the ${\rm
SU}(N_{TF})_L\times {\rm SU}(N_{TF})_R$ bulk gauge fields $A_L$ and
$A_R$, respectively. The value of vector and axial gauge fields,
defined as $V=(A_L+A_R)/\sqrt{2}$ and $A=(A_L-A_R)/\sqrt{2}\,$, at
the UV boundary couple to $J_{V\mu}^a$ and $J_{A\mu}^a$,
respectively.

The classical solution of the bulk field $X$ for $\Delta>2$ is given
as $X_0(z)=c_1\,z^{4-\Delta}+c_2\,z^{\Delta}$, where the constant
$c_1$ and $c_2$ are determined by the  boundary conditions. The
chiral condensate of techniquarks is (formally) defined as
\begin{equation}
\left<\bar q_L\,q_R\right>=i\left.\frac{\delta}{\delta
S}e^{iW[S]}\right|_{S=0}\equiv i\frac{\delta}{\delta
S}\left<e^{-i\int_x\left(\bar q_LSq_R+h.c.\right)}\right>_{S=0}\,,
\label{condensate}
\end{equation}
where $S$ is the source for the chiral condensate and the generating
functional is given by the AdS/CFT correspondence
\begin{equation}
W[S]=\int_x \left(-\frac{1}{4z^3}\,\partial_zX_0\,X_0^{\dagger}+h.c.
\right)_{z=\epsilon}\label{bulkaction}
\end{equation}
with the identification
$S(x)=\left.{2}\,X_0(x,z)/{z^{4-\Delta}}\right|_{z=\epsilon}\,$.
Since the source for the chiral condensate is the (bare) mass, we
find the UV boundary condition for $X_0$
\begin{equation}
\left.\frac{2}{z^{4-\Delta}}X_0\right|_{z=\epsilon}=M\,\quad{\rm
or}\quad c_1=\frac{1}{2}M\,.
\end{equation}
Then, $c_2=4\left<\bar q\,q\right>/{\Delta}$ in the chiral limit
($c_1\to 0$).

%\section{Peskin-Takeuchi S parameter}
The oblique S parameter is defined in terms of the two-point
function of the techniquark currents,
\begin{equation}
i\int_x e^{iq\cdot
x}\,\left<J_{\mu}(x)J_{\nu}(0)\right>=-(g_{\mu\nu}-{q_{\mu}q_{\nu}\over
q^2})\,\Pi(-q^2)
\end{equation}
as
\begin{equation}
S=4\pi\,\frac{\rm d}{ {\rm d}
q^2}\left[\Pi_V(-q^2)-\Pi_A(-q^2)\right]_{q^2=0},
\end{equation}
where $\Pi_V$ ($\Pi_A$) is the (axial) vector current correlator.
Since the source for the current is given by the bulk gauge fields,
the generating functional for the current correlation functions becomes
\begin{equation}
W[V,A]=-\frac{1}{2g_5^2}\int_x\left(\frac{1}{z}V_{\mu}^a\partial_z
V^{\mu a}+\frac{1}{z}A_{\mu}^a\partial_z A^{\mu
a}\right)_{z=\epsilon}\,, \label{generating}
\end{equation}
using the equations of motion. The UV boundary is identified as the
extended technicolor (ETC) scale, $\epsilon=1/\Lambda_{ETC}$. The IR
boundary is given by the technicolor scale, $z_m=1/\Lambda_{TC}$, at
which all techniquarks get mass and decouple.

The gauge fields in Eq.~(\ref{generating}) satisfy the bulk
equations of motion in unitary gauge,
\begin{eqnarray}
\left[\left(\partial^2-z\partial_z\frac{1}{z}\partial_z\right)\,
\eta_{\mu\nu}-\partial_{\mu}\partial_{\nu}\right]V^{\nu}&=&0\nonumber\\
\left[\left(\partial^2-z\partial_z\frac{1}{z}\partial_z+
\frac{g_5^2 X_0^2}{z^2}\right)\,
\eta_{\mu\nu}-\partial_{\mu}\partial_{\nu}\right]A^{\nu}&=&0,\label{gaugeeom}
\end{eqnarray}
where $\partial^2=\eta^{\mu\nu}\partial_{\mu}\partial_{\nu}$ and $X_0$
is a solution to the equation of motion for the bulk scalar field $X$,
together with suitable boundary conditions that satisfy
\begin{equation}
\sqrt{g}\,\left.{\rm Tr}\left(\delta A_{L\mu}F_L^{\mu z} +\delta
A_{R\mu}F_R^{\mu z}\right)\right|_{\epsilon}^{z_m}=0\,.
\end{equation}
For simplicity we choose $\delta A_{L\mu}=0=\delta A_{R\mu}$ at
$z=\epsilon$ and $F_L^{\mu z}=0=F_R^{\mu z}$ at $z=z_m$. Furthermore
we will work with the gauge $A_{Lz}=0=A_{Rz}$ to have
$\partial_zA_{L\mu}=0=\partial_zA_{R\mu}$ at $z=z_m$. We have
neglected the nonlinear terms for the gauge fields in
Eq.~(\ref{gaugeeom}), since we are interested only in the two-point
functions of the gauge fields, and this corresponds to keeping the
leading term in the large $N_{TC}$ expansion.

From Eqs.~(\ref{generating}) and (\ref{gaugeeom}) it is simple to
read off the two-point correlation functions of currents in momentum
space. We write the 4D Fourier transform of the vector and axial
gauge fields  as $V^\mu(q,z)=\left(g^{\mu\nu}-{q^\mu
q^\nu}/q^2\right) V(q,z), A^\mu(q,z)=\left(g^{\mu\nu}-{q^\mu
q^\nu}/q^2\right) A(q,z)+{q^\mu q^\nu}/q^2\,A(0,z)$ with
$V(q,\epsilon)=A(q,\epsilon)=1$ and $\partial_z V(q,z_m)=\partial_z
A(q,z_m)=0$, where $V(q,z)$ and $A(q,z)$ satisfy
\begin{eqnarray}
\left[z\partial_z\left({1\over z}\partial_z\right) +q^2\right] V(q,z)&=&0\nonumber\\
\left[z\partial_z\left({1\over z}\partial_z\right) +q^2-\frac{g_5^2
X_0^2}{z^2}\right] A(q,z)&=&0\,.
\end{eqnarray}
Then we have
\begin{equation}
\Pi_V(-q^2)=\left.\frac{\partial_z
V(q,z)}{z\,g_5^2}\right|_{z=\epsilon},
\Pi_A(-q^2)=\left.\frac{\partial_z
A(q,z)}{z\,g_5^2}\right|_{z=\epsilon}\,.
\end{equation}

The solution for $V(q,z)$ is given as, with
$\left|q\right|=\sqrt{\left|q^2\right|}$,
\begin{equation}
V(q,z)=a_1\left|q\right|z\,
Y_1(\left|q\right|z)+a_2\left|q\right|z\, J_1(\left|q\right|z)\,,
\end{equation}
where $J_1$ and $Y_1$ are the first order Bessel and Neumann
functions, respectively. The constant $a_1$ and $a_2$ are to be
fixed by the boundary conditions, $V(q,\epsilon)=1$ and
$\partial_zV(q,z_m)=0$. From this we have
\begin{equation}
\Pi_V(-q^2)=
\frac{|q|}{g_5^2\epsilon}\frac{J_0(|q|z_m)Y_0(|q|\epsilon)-
Y_0(|q|z_m)J_0(|q|\epsilon)}{
J_0(|q|z_m)Y_1(|q|\epsilon)-Y_0(|q|z_m)J_1(|q|\epsilon)}\,,
\label{vector}
\end{equation}
which becomes $q^2\,\ln\left(z_m/\epsilon\right)/g_5^2$ as $q^2\to
0$. Comparing it also with the perturbative calculation at large
momentum, $-q^2\to\infty$\,,
\begin{equation}
\Pi_V(-q^2)=\frac{d_R}{24\pi^2}\,q^2\ln\left(-q^2\right)+\cdots\,,
\end{equation}
we match $g_5^2={12\pi^2}/{d_R}$, where $d_R=N_{TC}$ and
${N_{TC}(N_{TC}+1) / 2}$ are the dimensions for fundamental and
symmetric second-rank tensor representations, respectively. We
resort, however, to numerical analysis for $A(q,z)$ and
correspondingly for the S parameter.

%\section{Walking}
In technicolor theory quarks and leptons get mass through coupling
to techniquarks by an extended technicolor (ETC)
interaction~\cite{Hill:2002ap}, whose scale has both lower and upper
bounds, coming from the constraint to generate the observed mass,
while suppressing the flavor-changing neutral current processes. The
QCD-like technicolor fails to generate an ETC scale that satisfies
the constraint. However, if the $\beta$-function of the technicolor
theory has a quasi IR fixed point, $\beta(\alpha_*)=0$, the coupling
is almost constant near the fixed point and thus the strength of the
bilinear operator gets enhanced at low energy as
\begin{equation}
\left.\bar q q\right|_{\Lambda_{\rm
ETC}}\simeq\left(\frac{\Lambda_{\rm ETC}}{\Lambda_{\rm
TC}}\right)^{\gamma_m}\,\left.\bar q q\right|_{\Lambda_{\rm TC}},
\end{equation}
where $\gamma_m$ is the anomalous dimension of the bilinear operator
near the fixed point. The Schwinger-Dyson analysis shows
$\gamma_m=1$ in the phase where the chiral symmetry is
broken~\cite{Cohen:1988sq,Hong:1989zz}. Therefore, in the
technicolor theories with a quasi IR fixed point, which will be
called  conformal technicolor theories in short, a large enough mass
for quarks and leptons is possible even with a large hierarchy
between $\Lambda_{\rm ETC}$ and $\Lambda_{\rm TC}$.

The holographic dual of such conformal technicolor is very different
from that of QCD-like technicolor. The mass of the bulk scalar
becomes $m_5^2=-4$, saturating the Breitenlohner-Freedman bound,
since the scaling dimension of $\bar qq$ in the conformal
technicolor is 2 instead of 3. The classical solution becomes
$X_0(z)=c_1\,z^2+c_2\,z^2\ln(z/\epsilon)$.  The UV boundary
condition fixes the constant $c_1=0$ in the chiral limit and by
AdS/CFT correspondence $c_2=\left<\bar q \,q\right>/4$.

To estimate the S parameter of conformal technicolor theories using
holography, we solve
\begin{equation}
\left[ z\partial_z
\left(\frac{1}{z}\partial_z\right)+q^2-\frac{g_5^2X_0^2}{z^2}\right]A(q,z)=0\,,
\label{axial}
\end{equation}
with the boundary conditions $A(q,\epsilon)=1$ and $\partial_z
A(q,z_m)=0$. Since Eq.~(\ref{axial}) is linear in $A(q,z)$ and
$\Pi_A(q^2)$ is related to $\partial_z A(q,\epsilon)$, it is
convenient to work instead with $T(q,z)\equiv\partial_z\ln\,A(q,z)$
with a {\it single} boundary condition, $T(q,z_m)=0$. Then,
$\Pi_A(-q^2)$ simply becomes $T(q,\epsilon)/(g_5^2 \epsilon)$.
Expanding $T(q,z)=T^{(0)}(z)+q^2T^{(1)}(z)+\cdots$, we rewrite
Eq.~(\ref{axial}) as
\begin{eqnarray}
z\partial_z\left(\frac{1}{z}T^{(0)}\right)+\left(T^{(0)}\right)^2&=&
\frac{g_5^2X_0^2}{z^2}\,,
\label{zero}\\
z\partial_z\left(\frac{1}{z}T^{(1)}\right)+2T^{(0)}T^{(1)}&=&-1\,,
\label{one}
\end{eqnarray}
and so on. Solving Eq.~(\ref{one}), we get
\begin{equation}
T^{(1)}(z)=-z\,\int_{z_m}^z\frac{{\rm d}z^{\prime}}{z^{\prime}}\,
\exp\left[{2\int_z^{z^{\prime}}{\rm
d}\omega\,T^{(0)}(\omega)}\right]\,.
\end{equation}
Since  $g_5^2\,
\Pi_V(-q^2)=\ln\left(z_m/\epsilon\right)\,q^2+O(q^4)$ for small
$q^2$, the S parameter becomes
\begin{equation}
S={4\pi\over g_5^2}\int_{\epsilon}^{z_m}\frac{{\rm d}z^{\prime}}{z^{\prime}}\,
\left[1-e^{2\int_{\epsilon}^{z^{\prime}}{\rm
d}\omega\,T^{(0)}(\omega)}\right]\,.\label{S}
\end{equation}
To calculate the S parameter, we solve $T^{(0)}$ in (\ref{zero})
numerically with the boundary conditions $T^{(0)}(z_m)=0$ and
\begin{equation}
-\Pi_A(0)=-\left.\frac{1}{g_5^2}\frac{T^{(0)}(z)}{z}\right|_{\epsilon}=F_T^2\,,
\end{equation}
where the technipion decay constant $F_T=246\sqrt{2/N_{TF}}~{\rm
GeV}$.

We solve Eq.~(\ref{zero}) for given boundary conditions,
$T^{(0)}(z_m)=0$ and $T^{(0)}(\epsilon)/\epsilon=-g_5^2F_T^2$, using
two different bulk scalar fields: $X_0=c_Wz^2\ln(z/\epsilon)$ for
the technicolor with walking behavior (namely for the
techni-orientifold and the walking technicolor) and
$X_0=c_Q\left(z^3-z\,\epsilon^2\right)$ for the QCD-like
technicolor. Since
$\left.\partial_zT^{(0)}\right|_{z_m}=g_5^2\,{X_0^2}/{z_m^2}$, there
is a unique value for $c_{W,Q}$ or $\left<\bar q q\right>$ which
allows a solution for $T^{(0)}$. In the holographic dual therefore
$F_T$ determines $\left<\bar q q\right>$ or vice versa.

In the ladder approximation the fermion self energy  becomes
$\Sigma(p)={\kappa}/{p}\,$ for a large Euclidean momentum, $p\to
\infty$, which then gives $\left<\bar q q\right>=\Lambda\,\kappa\,$
at a scale $\Lambda$ , if compared with the operator product
expansion~\cite{Cohen:1988sq}. By analyzing the gap equation in the
ladder approximation, one further finds that
$\kappa=\Sigma^2(0)$~\cite{Hong:1989zz}. Since the dynamical mass of
techniquarks $\Sigma(0)\simeq\Lambda_{TC}$, $\left<\bar q
q\right>\simeq1/z_m^3$ at $\Lambda_{TC}$ or $c_Wz_m^3\simeq1/4$.

Our results for S parameter depend only on two dimensionless
parameters $\epsilon/z_m$ and $F_T z_m$. For the former, we take
$1/300=\Lambda_{TC}/\Lambda_{ETC}$, while the latter must be
estimated by other means. If we  use the ladder approximation value
$c_Wz_m^3=1/4$, we obtain $F_Tz_m\simeq0.86/g_5$ by solving
Eq.~(\ref{zero})\,. The corresponding S parameters are listed in the
first row of the table~I. If the chiral perturbation theory is valid
up to the scale $m_\rho$, which is the first pole of $\Pi_V$ in
Eq.~(\ref{vector}),  we have $4\pi F_T\simeq m_\rho$ or $F_T\simeq
0.19/z_m$. Finally, we include the QCD value, $m_\rho/F_T\simeq
8.85$ or $F_T\simeq0.29/z_m$, for comparison. (For a strict comparison
one should take into account both $F_T$ and $z_m$ depend on $N_{TC}$
and the representation of techniquarks.)
\begin{center}
\begin{table}[htb]
\begin{tabular}{|c|c|c|c|c|c|c|}
\hline $F_T z_m$ & $N\!\!=2,S$ & $N\!\!=2,F$& $N\!\!=3,S$ &
$N\!\!=3,F$
& $N\!\!=4,S$ & $N\!\!=4,F$ \\
\hline
$0.86/g_5$ & 0.086 & 0.057 & 0.17 & 0.086 & 0.29 & 0.12  \\
%$1/4\pi$ & 0.031 & 0.031 & 0.031 & 0.031 & 0.031 & 0.031 \\
0.19 & 0.15 & 0.14 & 0.17 & 0.15 & 0.17 &  0.16\\
0.29 & 0.28 & 0.22 & 0.34 & 0.26 & 0.37 & 0.31 \\
\hline
\end{tabular}
 \caption{The S parameter for conformal technicolor with
techniquarks in the symmetric second-rank tensor (S) and fundamental
(F) representations.}\label{tablet}
\end{table}
\end{center}

For the QCD-like technicolor with $F_T=0.29/z_m$ we find
$S=0.24,0.30,0.34$ for $N_{TC}=2,3,4$, respectively, which agrees
well with the estimate by rescaling the QCD
data~\cite{Peskin:1990zt}. In Table~\ref{tablet}, we list our
numerical results for other technicolor models. Several comments are
in order. First, we note that all the S parameters we calculated are
positive. In fact one can prove the $S$ parameter is always positive
in the holographic dual of any models for dynamical electroweak
symmetry breaking. By examining Eq.~(\ref{zero}) one can show that
$T^{(0)}$ is always negative for any value of $X_0$. Suppose
$T^{(0)}(z)$ has one more zeros for $z<z_m$. Since the right-hand
side of Eq.~(\ref{zero}) is always positive, the slope of
$T^{(0)}/z$ has to be positive at those zeros, which is however
impossible for a continuous function. Therefore $T^{(0)}$ must have
no zeros and thus negative for $z<z_m$. The $S$ parameter in
Eq.~(\ref{S}) is hence always positive.

\begin{figure}
%\vskip 0.2in
\epsfxsize=3.5in \centerline{\epsffile{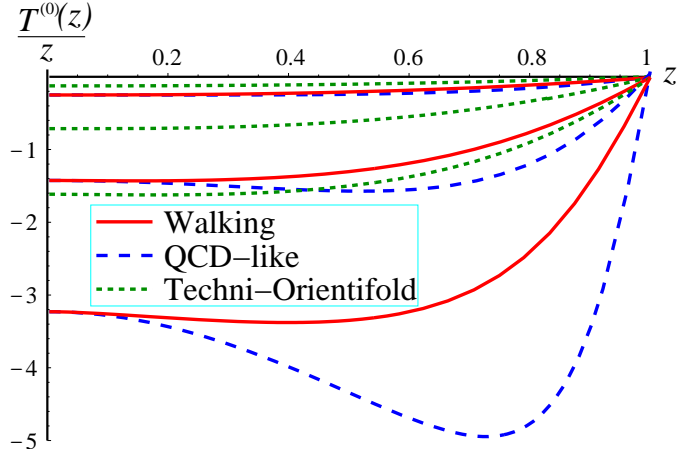}} \caption{The
profiles  for $N_{TC}=3$ with various
$T^{(0)}(\epsilon)/\epsilon=-g_5^2F_T^2$. }
 \label{fig2}
\end{figure}

Second, we find that the S parameter is reduced about 10-20\% by
walking when $F_T$ and $d_R$ are same, which agrees with the
Weinberg sum rule~\cite{Sundrum:1991rf}. This can be seen easily,
since  the profile $-T^{(0)}(z)$ is always smaller for the walking
technicolor than for the QCD-like technicolor with the same $F_T$
and $d_R$ and so is the S parameter. (See Fig.~\ref{fig2}, where we
take $z_m=1$.) Suppose $\partial_z\left(T^{(0)}/z\right)=0$ at
$z=z_*>\epsilon$. Then, $\partial_z\left(T^{(0)}/z\right)\simeq
h\,\left(z-z_*\right)$ near $z_*$. Expanding Eq.~(\ref{zero}) around
$z=z_*$ for the walking case,  we find $h>0$ and
$z_*=\epsilon\,e^{h/(2g_5^2c_W^2)}\,$. The slope of $T^{(0)}/z$
therefore vanishes only at a point very close to the UV boundary.
$\partial_z\left(T^{(0)}/z\right)$ is positive for $0<z\le z_m$ if
we take $\epsilon\to0$. On the other hand for the QCD-like
technicolor, the slope vanishes whenever, taking $\epsilon\to0$,
\begin{equation}
\frac{1}{z}\,T^{(0)}(z)=-g_5c_Q\,z\,. \label{slope}
\end{equation}
Since $T^{(0)}(z)/z$ is finite at $z=0$ but vanishes at $z_m$ while
the right-hand side of Eq.~(\ref{slope}) is monotonically
decreasing, there must exist a solution $z_*$ to Eq.~(\ref{slope}),
remaining finite when $\epsilon\to0$. We now note that the bulk
fields $X_0$ for the walking technicolor and the QCD-like
technicolor are monotonically increasing functions and meet together
only at two points $z=\epsilon$ and $z=z_0$. If $z_0>z_m$, the slope
of $T^{(0)}/z$ at $z_m$ is bigger for the walking and the
$T^{(0)}/z$ profiles must meet at a point between $z_*$ of the
QCD-like technicolor and $z_m$, which is impossible however because
then the slope has to be smaller for the walking case at the point
though its $X_0$ is bigger. Therefore $z_0<z_m$ or the slope of
$T^{(0)}/z$ at $z_m$ has to be smaller for the walking technicolor
and thus its profile $-T^{(0)}(z)$ is always smaller than that of
QCD-like technicolor. The S parameter is therefore somewhat reduced
by walking. However, we expect further reduction in the S parameter
for the walking case, since $F_T/\Lambda_{TC}$ may be quite small in
the walking technicolor~\cite{Harada:2005ru}.

Finally, we find that a slight change in $F_T$ results in a
substantial change in the S parameter. As shown in Fig.~\ref{fig2},
the area sustained by $-T^{(0)}/z$ mainly depends on its value at
$z=\epsilon$.  If $F_T$ is much smaller than $\Lambda_{TC}$, the S
parameter gets reduced substantially.

%\section{conclusion}
To conclude, we have analyzed the oblique S parameter in the
holographic dual of technicolor theories. The AdS/CFT correspondence
allows us to investigate the general aspects of the S parameter and
also to obtain its numerical values precisely, which therefore
removes a major hurdle for technicolor theories. We have also shown
that the S parameter is strictly positive in the holographic dual
models of technicolor and is reduced at least 10-20\% by walking. We
predict $F_T/\Lambda_{TC}=0.86/g_5$ in terms of the 5D gauge
coupling in the ladder approximation, which results in small S
parameters. For the technicolor models with $N_{TF}>2$ the S
parameter will increase but not much since $F_T$ decreases as
$1/\sqrt{N_{TF}}$.

%\vfill \eject

%\acknowledgments
The work of D.~K.~H. is supported by Korea Research Foundation Grant
funded by Korean Government (MOEHRD) (KRF-2005-015-C00108) and
H.~U.~Y is supported by grant No. R01-2003-000-10391-0 from the
Basic Research Program of the Korea Science \& Engineering
Foundation.

\end{document}